**Ubiquitous Antiparallel Domains in 2D Hexagonal Boron Nitride Uncovered by Interferometric Nonlinear Optical Imaging**

*Yeri Lee, Juseung Oh, Kyung Yeol Ma, Seung Jin Lee, Eui Young Jung, Yani Wang, Kenji Watanabe, Takashi Taniguchi, Hailin Peng, Hiroki Ago, Ki Kang Kim, Hyeon Suk Shin and Sunmin Ryu**

Y. Lee, J. Oh, S. Ryu

Department of Chemistry, Pohang University of Science and Technology (POSTECH), Pohang, Gyeongbuk 37673, Republic of Korea.

E-mail: sunryu@postech.ac.kr

K. Y. Ma

Department of Chemistry, Ulsan National Institute of Science and Technology (UNIST), Ulsan 44919, Republic of Korea.

Research Laboratory of Electronics, Massachusetts Institute of Technology, Cambridge, MA 02139, USA

S. J. Lee

Department of Energy Science, Sungkyunkwan University (SKKU), Suwon 16419, Republic of Korea.

E. Y. Jung, H. Ago

Interdisciplinary Graduate School of Engineering Sciences, Kyushu University, Fukuoka 816-8580, Japan.

Y. Wang, H. Peng

Center for Nanochemistry, Beijing Science and Engineering Center for Nanocarbons, Beijing National Laboratory for Molecular Sciences, College of Chemistry and Molecular Engineering, Peking University, Beijing 100871, China.

K. Watanabe

Research Center for Electronic and Optical Materials, National Institute for Materials Science, 1-1 Namiki, Tsukuba 305-0044, Japan.

T. Taniguchi

Research Center for Materials Nanoarchitectonics, National Institute for Materials Science, 1-1 Namiki, Tsukuba 305-0044, Japan.

K. K. Kim

Department of Energy Science, Sungkyunkwan University (SKKU), Suwon 16419, Republic of Korea.

Department of Physics, Sungkyunkwan University (SKKU), Suwon 16419, Republic of Korea.

H. S. Shin

Department of Chemistry, Ulsan National Institute of Science and Technology, Ulsan 44919, Republic of Korea.

Department of Energy Science, Sungkyunkwan University (SKKU), Suwon 16419, Republic of Korea.

Center for 2D Quantum Heterostructures, Institute of Basic Science (IBS), Sungkyunkwan University (SKKU), Suwon 16419, Republic of Korea.

**Abstract**

Hexagonal boron nitride (hBN) supports a wide range of two-dimensional (2D) technologies, yet assessing its crystalline quality over large areas remains a fundamental challenge. Both antiparallel domains, an intrinsic outcome of epitaxy on high-symmetry substrates, and associated structural defects have long evaded optical detection. Here, we show that interferometric second-harmonic generation (SHG) imaging provides a powerful, non-destructive probe of lattice orientation and structural integrity in 2D hBN grown by chemical vapor deposition. This approach reveals the ubiquitous formation of antiparallel domains and quantifies their impact on crystalline order. SHG intensity also emerges as a direct optical metric of domain disorder, spanning three orders of magnitude across films produced by ten different growth routes. Correlation with Raman spectroscopy establishes a unified framework for evaluating crystalline quality. Beyond hBN, this method offers a high-throughput route to wide-area structural imaging in various non-centrosymmetric materials, advancing their deployment in electronics, photonics, and quantum technologies.

## 1. Introduction

Since the isolation of graphene [1], enormous efforts have been devoted to growing high-quality two-dimensional (2D) crystals such as graphene [2], transition metal dichalcogenides (TMDs) [3] and hexagonal boron nitride (hBN) [4], owing to their remarkable physical properties and broad technological potential. Among them, hBN has attracted particular interest for its wide bandgap [5, 6], high thermal conductivity [7], chemical stability [8], and mechanical robustness [9], making it indispensable as a dielectric in van der Waals heterostructures and promising for diverse electronic [10, 11], optoelectronic [12], and energy applications [13]. Chemical vapor deposition (CVD) on crystalline metal substrates with high-symmetry facets has been a widely adopted route for synthesizing 2D crystals with large domain sizes, monolayer controllability, and high crystallinity [14]. For graphene, this strategy has yielded meter-scale single crystals on Cu(111) [15]. For hBN, however, achieving comparable crystalline perfection over a large area has proven considerably more difficult, although a range of metals, including Cu [16-18], Ni [19, 20], Fe-Ni [21], Au [22], Pt [23], Co [24], and Ru [25], have been investigated as epitaxial substrates. The challenge arises partly from its lower, threefold rotational symmetry, which allows two crystallographically equivalent but antiparallel orientations to nucleate on sixfold-symmetric metal surfaces such as Cu(111) [17, 26]. When these domains merge, grain boundaries form and compromise optical and electronic properties [26]. To address this, high-index vicinal facets and modified growth strategies have been explored, including for hBN [17, 27] and $MoS_2$ [26], yet the field still faces an urgent need for a non-destructive and efficient structural probe that can evaluate large-area samples and capture domain disorders, including antiparallel domains spanning multiple length scales.

Conventional atomic-resolution methods such as transmission electron microscopy (TEM) [16-18] and scanning tunneling microscopy (STM) [17, 18] provide atomic-scale structural information but are limited in sampling area and require demanding sample preparation. Low-energy electron diffraction (LEED) [17, 22] offers millimeter-scale probing but cannot distinguish antiparallel domains. Raman spectroscopy has been widely used due to its structure-specificity and non-destructive nature [20, 28], yet it cannot resolve orientational domains, let alone the polarities of non-centrosymmetric crystals such as hBN and TMDs. The vibrational method does not provide spectroscopic features for structural disorder of hBN other than the Raman linewidth of the $E_{2g}$ mode at 1366 $cm^{-1}$, as it does for graphitic materials exhibiting the disorder-induced D band [29]. Whereas photoluminescence spectroscopy reveals emissions originating from atom-scale defects [30], it is not specific to domain boundaries and is limited by the technical requirements of excitation sources and detection optics specialized for the UV range. Nonlinear optical spectroscopy, such as second-harmonic generation (SHG), overcomes these limitations when combined with phase-resolving heterodyne detection [31-34]. Most of all, SHG is generally sensitive to crystallographic orientation because of the lower symmetry of the second-order electric susceptibility [35]. The coherent nature of SHG signals renders the method highly sensitive to the presence of antiparallel domains [35, 36] and stacking order [35]. Heterodyne SHG spectroscopy, also measuring the relative phase of SHG signals [31], enabled the determination of absolute molecular orientation [32], stimulated SHG [37], identification of electrical double layers [38] and a comprehensive description of TMD heterostructures [33, 34, 39]. A similar approach can be applied to sum frequency generation [40] and degenerative optical amplification [41]. Recently, Mueller et al. used resonant sum-frequency generation for crystallographic imaging of hBN [42]. More broadly, heterodyne detection has been utilized in other coherent nonlinear optical modalities, such as coherent anti-Stokes Raman scattering (CARS) microscopy, to suppress nonresonant background signals [43, 44]. Despite this potential and recent advances, extremely low SHG signals [42] and disorder across multiple length scales in hBN continue to hinder scalable optical characterization and underscore the lack of reliable, large-area metrics for crystalline quality.

Here we present interferometric SHG polarimetry, which reveals antiparallel domains spanning multiple length scales. By combining polarization-resolved SHG with interferometric detection, we optically resolve antiparallel domains and their boundaries in CVD-grown hBN, uncovering their ubiquitous presence on high-symmetry substrates. A mixed-domain model reveals that SHG intensity is a sensitive optical metric of crystalline order, varying over a few orders of magnitude depending on growth method. By benchmarking against mechanically exfoliated hBN, we establish correlations among SHG response, Raman intensity, linewidth, and orientational disorder, thus providing a unified framework for structural assessment. The demonstrated methodology is broadly applicable to TMDs and other non-centrosymmetric materials, offering a scalable approach to structural imaging and quality assessment.

## 2. Results and Discussion

### 2.1. Structure-sensitivity of coherent SHG signals.

Generated by a two-photon parametric process that does not alter the quantum state of the system (**Figure 1a**), SHG signals exhibit several unique characteristics compared with spontaneous Raman scattering. First, the second-order electric susceptibility governing SHG is a third-rank tensor, which generally leads to a less symmetric response than the linear counterpart [45]. Consequently, the nonlinear optical signals can be anisotropic even when the linear response is isotropic, as is the case for hBN belonging to the $D_{3h}^1$ space group [35] (Figure 1a). Second, SHG signals are coherent and formed by the superposition of all second-harmonic (SH) electric fields within the probing volume [33, 35, 46]. Thus, SHG intensity is proportional to the square of the sample thickness when constructive interference is maintained, as in AB-stacked bilayer (2L) hBN [47] (Figure 1b). In contrast, SHG intensity vanishes, when half of the sample is aligned antiparallel to the other half, as in AA'-stacked 2L hBN (Figure 1b) [35]. A similar destructive interference occurs when the excitation beam is focused at the boundary between two in-plane antiparallel domains (Figure 1c, top). Notably, zero SHG intensity has been the hallmark of domain boundaries between antiparallel TMDs spanning micrometer-scale lengths [36, 48]. An even more intriguing case arises when the domain size is much smaller than the beam spot. If the focal area contains multiple domains (Figure 1c, bottom), the SHG intensity depends on the relative population of the two antiparallel orientations, becoming zero when both are equally represented. This feature makes SHG particularly sensitive to the presence of antiparallel domains. Finally, the coherent nature of SHG allows discrimination between fields generated in antiparallel domains. As illustrated in the top panel of Figure 1c, the two antiparallel domains produce SH fields that are out of phase with each other. The phase difference can be measured using interferometric detection with a local oscillator field [33, 39]. These characteristics render interferometric SHG spectroscopy and imaging a powerful tool for orientational polarimetry.

### 2.2. Widely varying SHG response of CVD-grown hBN.

We show that the SHG signals of 2D hBN vary by up to three orders of magnitude among samples synthesized by different methods, unlike Raman scattering. For comparative studies for linear and nonlinear spectroscopies, ten types of large-area single or few-layer hBN films grown under ten different CVD conditions were prepared on Si substrates with an 85 nm-thick $SiO_2$ epilayer (Table S1; also see Methods for details). **Figure 2a** and 2c show the optical micrographs of a representative 3L hBN sample (denoted as $3L_{U1}$, where U1 specifies the growth condition in Table S1) and a reference monolayer ($1L_{ME}$), respectively. The latter was mechanically exfoliated (ME) from high-quality bulk crystals grown under high pressure [49]. For $3L_{U1}$, we selected an area with tears and folds (Figure 2a) as intact regions showed almost no optical and topographic contrast. The apparent step height across the folded area in the atomic force microscopy (AFM) height image (Figure 2b) was ~2.5 nm, larger than the 3L thickness verified by cross-sectional TEM [20]. This should therefore be interpreted as an apparent AFM height, likely overestimated due to interfacial gaps filled with foreign materials introduced during transfer and folding [50, 51]. By contrast, the step height of $1L_{ME}$ was ~0.4 nm (Figure 2d), consistent with the interlayer spacing of hBN crystals [52] and indicating a cleaner interface.

Optical characterization using reference 2D hBN samples requires reliable thickness determination, which is less straightforward [53] than for graphene [54] or $MoS_2$ [55]. As shown in Figure 2e, the optical contrast (OC) [56] did not resolve monolayer thickness in hBN because of its negligible visible absorption [53]. Here, OC was defined as $(R - R_o)/R_o$, where R and $R_o$ are the red-channel intensities of sample and bare substrate regions in optical micrographs such as Figure 2c. The OC increased sub-linearly with thickness up to six layers, rather than linearly, as also observed in TMDs [57]. This behavior contrasts with earlier reports on hBN, where OC varied linearly up to three layers [53]. Although SHG intensity (Figure 2e) facilitated thickness identification owing to its even-odd alternation [35], some samples with large OC spread could not be unambiguously assigned. To complement this, we also measured Si-referenced Raman intensity of the $E_{2g}$ mode at 1366 $cm^{-1}$, characteristic of hBN [58]. Normalization to the substrate Raman signals minimize extraneous variations, allowing monolayer-level resolution (red circles in Figure 2e).

Representative unpolarized SHG spectra of $3L_{U1}$ and $1L_{ME}$ are shown in Figure 2f. Both exhibit quadratic dependence on the power of the fundamental beam (inset), but strikingly, $3L_{U1}$ yielded ~500 times weaker SHG than the exfoliated reference. As described below, all CVD samples exhibited wide spot-to-spot and sample-to-sample variations, unlike exfoliated references. On average, normalized SHG intensities ranged from ~0.1% to 100% of the $1L_{ME}$. Some types of crystalline disorders causing such variations can be probed by polarized SHG spectroscopy (Figure 2g). For odd-layer hBN ($D_{3h}^1$ space group), the polarized SHG intensity follows sixfold azimuthal symmetry: $I_{2\omega}^{\parallel}$ ($I_{2\omega}^{\perp}$) $\propto \cos^2 3\theta$ ($\sin^2 3\theta$), where $I_{2\omega}^{\parallel}$ ($I_{2\omega}^{\perp}$) denotes SHG signals polarized parallel (perpendicular) to the incident polarization [33, 35]. In Figure 2h, $I_{2\omega}^{\parallel}$ of $3L_{U1}$ showed 60° periodic modulation but with symmetry strongly distorted compared with that of $1L_{ME}$ (Figure 1a, bottom). Similar asymmetry appeared in other CVD samples (Figure S1). While lattice distortion, as in strained $MoS_2$ [59], may contribute, spatial inhomogeneity was identified as the major cause: birefringence of the half-wave plate caused the focal spot to wobble along a circle (diameter of 2.4 μm) during angle scans (Methods).

When replaced with another type of half-wave plate (wobble diameter 0.4 μm), the asymmetry was greatly reduced (Figure S2).

Despite the widely varying SHG response, most CVD samples exhibited a prominent $E_{2g}$ Raman mode (Figure 2i). Their peak position, 1368.5 ± 2.6 $cm^{-1}$, matched that of the $1L_{ME}$, which is susceptible to substrate-induced strain and subject to peak shifts of a few $cm^{-1}$ . The Raman intensity varied within a factor of four compared with the reference (quantitative analysis below). Except for one type of sample ($1L_{C1}$), which showed notable broadening, the FWHM of $E_{2g}$ ranged from 13 to 24 $cm^{-1}$, consistent with literature [28]. This fact suggests the process of extended disorders because Raman linewidth, often linked to crystallinity, is relatively insensitive to the types of defects detectable by cathodoluminescence [28]. Assuming Raman intensity scales with the number of crystalline unit cells, a fourfold decrease in Raman intensity for $3L_{U1}$ would correspond to a 16-fold decrease in SHG intensity, given its quadratic dependence. The actual ~500-fold suppression (Figure 2f) thus points to additional structural disorders strongly attenuating SHG, despite a large fraction of crystalline units being preserved.

**2.3. Domain imaging by polarimetric SHG**

The distorted SHG polar graphs and spot-to-spot intensity variations suggested significant spatial inhomogeneity, which may originate from point defects or extended structural disorders. To address this, we performed dual-polarization SHG imaging (**Figure 3a**) and mapped crystallographic orientation (θ) using the trigonometric relation: $\theta = \frac{1}{3}\tan^{-1}\sqrt{\frac{I_{2\omega}^{\perp}}{I_{2\omega}^{\parallel}}}$ , where θ varies from 0 to 30° (Figure 2g) [35, 36]. It should be noted that the relation cannot differentiate antiparallel domains. As shown below, each type of hBN samples exhibited a distinctive distribution of orientational domains. Figure 3b and 3c show the images of $I_{2\omega}^{\parallel}$ and $I_{2\omega}^{\perp}$ simultaneously acquired during raster scanning of a few-layer polycrystalline sample ($fL_{U3}$) grown on Ni substrates. Whereas $I_{2\omega}^{\parallel}$ and $I_{2\omega}^{\perp}$ showed complementary intensity variations across domain boundaries (dotted lines), their sum ($I_{2\omega}^{\parallel+\perp}$) was more or less uniform (Figure 3d). However, ~60% spot-to-spot variation in $I_{2\omega}^{\parallel+\perp}$ (Figure 3e) suggests the presence of structural disorder. By contrast, the variation in $1L_{ME}$ was only ~1.3% (orange line in Figure 3e and Figure S3e). Note that $I_{2\omega}^{\parallel+\perp}$ is orientation-independent (Figure S4). From the global variation of $I_{2\omega}^{\parallel}$ and $I_{2\omega}^{\perp}$, we infer that the upper right (lower left) region of Figure 3c generates mostly horizontally (vertically) polarized SH fields. The θ-image in Figure 3f indeed revealed two dominant orientational domains. The frequency graph in Figure 3g showed that their orientations, defined in Figure 2g and 3f (inset), correspond to 9° and 24° relative to the y-axis.

Notably, the frequency graph of θ-image displayed a broad spread: the FWHM of the two components was ~4.6°, ~35 times larger than that of $1L_{ME}$ (0.13° in Figure S3f). As discussed below, the large orientational spread arises from lattice disorder.

Unique SHG intensity patterns were also observed in 1L samples continuously grown by Cu catalysts (denoted as $1L_{S1}$). The $I_{2\omega}^{\parallel+\perp}$ (Figure 3h) and polarized intensity (Figure S5) images indicate that randomly shaped domains spanning ~10 μm are separated by meandering, thick boundaries that exhibit weak SHG signals (Figure 3i). Notably, the frequency graph of θ-image in Figure 3j revealed that the entire region, including the boundaries, is nearly unidirectional. Figure 3k showed two orientational components differing by only 0.9°. The minor population centered at 10.4° corresponded to the stripes in Figure 3j (red arrows), attributable to structural disorder caused by long-range undulations in Cu catalysts [60]. Unlike $fL_{U3}$ (Figure 3g), the two components in Figure 3k exhibited a reduced width of ~0.9°. As shown in the inset of Figure 3k, the orientational spread within individual domains (e.g., white circle in Figure 3h) was slightly larger than that of $1L_{ME}$. These results indicate that $1L_{S1}$ consists of numerous high-quality crystals stitched together in an apparently unidirectional manner. As shown below, however, some are antiparallel to others (Figure 1c), forming characteristic boundaries.

Figure 3l presents the $I_{2\omega}^{\parallel+\perp}$ image of a few-layer hBN grown on polycrystalline Fe-Ni alloy catalysts (denoted as $fL_{K1}$). The image shows that the continuous hBN film is composed of numerous crystallites, each with relatively uniform intensity (see Figure S5 for polarized intensity images). The broad intensity distribution in Figure 3m, reflecting heterogeneity in thickness and stacking order, is distinct from that in Figure 3i, which arises mainly from domain mixing within the laser focal spot. The θ-image in Figure 3n revealed more distinguishable domains than the intensity image, with typical sizes on the order of 10 μm, comparable to $1L_{S1}$ (Figure 3h). However, unlike $1L_{S1}$, the domains of $fL_{K1}$ were randomly oriented (Figure 3o), consistent with the polycrystalline nature of the catalysts (Table S1). Importantly, the orientational spread within each domain (insets in Figure 3k and 3o) was narrow and comparable to that of $1L_{ME}$. Note that the entire distribution in Figure 3o summarizes the film-wide orientational diversity, whereas the inset shows a single-domain orientational spread relevant in assessing the structural quality of each domain.

**2.4. SHG interferometry for antiparallel domains.**

Whereas SHG intensity and orientational imaging reveal substantial spatial variations in CVD-grown hBN (Figure 2 and 3), intensity polarimetry alone cannot distinguish antiparallel domains, which led us to exploit interferometric SHG spectroscopy [31-34]. As depicted in **Figure 4a**, two antiparallel domains (e.g., $\theta = 0°$ and $60°$ in Figure 4a) yield oppositely-facing SH fields, which result in contrasting interferences when probed with an external SH local oscillator (LO) [61]. We implemented in-line spectral phase interferometry [33, 39] (Figure 4a) using an α-quartz crystal to generate LO fields (Methods). In Figure 4b, the interferograms of $1L_{ME}$ reference sample at $\theta = 0°$ and $60°$ exhibited a half-period difference ($\pi$ in phase), which is consistent with expectations and prior reports[32].

Using interferometric detection, we demonstrate the presence of antiparallel domains in $1L_{S1}$, which otherwise exhibit Raman and SHG features consistent with high crystallinity. In Figure 4d and 4e, pixels with SHG intensity < 30% of the average were assigned as boundaries and colored in black (see Figure S6 for intensity and raw orientation images). While the overall SHG intensity ($I_{2\omega}^{\parallel+\perp}$) was comparable to the reference, boundaries exhibited reduced SHG signals despite orientations nearly matching surrounding domains (Figure S6). To probe the structure of boundary regions, we performed line scans along the white dashed line in Figure 4d. The SHG phase ($\varphi_{SHG}$) profile in Figure 4f revealed that two boundary regions (α and γ) were 180° out of phase with the surrounding regions (β and δ), showing that the former is antiparallel to the latter. A high-resolution image (Figure 4e) further revealed that the apparent boundaries α and γ in Figure 4d are composed of smaller subdomains (α' and γ') with β-like phase values and thinner boundary areas. Then, it is likely that any boundary region consists of many sub-domains belonging to either of the two domain types depicted in Figure 1c: its SHG signals are reduced via destructive interference between the two groups, and its SHG phase is determined by the majority group. These results indicate that two neighboring domains (e.g., α' and β') separated by a closed boundary are antiparallel, supporting a simple growth model in which antiparallel domains nucleate independently and merge to form boundaries. As shown in Figure S7, the same conclusion was drawn from a phase line profile obtained from the orange dashed line in Figure 4d. As shown in Figure 4c, the phase inversion was verified across 37 domain boundaries in two independently grown monolayer samples ($1L_{S1}$ and $1L_{K3}$), in which similar boundary-associated phase structures were observed. We also note that the boundary spots exhibited distinctive Raman behavior (Figure 4f): although their $I_{2\omega}^{\parallel+\perp}$ dropped below 20% of the reference, their Raman intensity ($I_R$) remained comparable to (α) or even exceeded (γ) that of β or δ. The ~50% Raman enhancement at γ may indicate the presence of additional hBN materials within the focal area, such as wrinkles, folds or multilayer patches [62], although we do not assign a unique structural origin from the present data. Raman linewidth (Γ) was on average ~50% broader than the reference, with even larger broadening at β.

We also applied the interferometric SHG to samples with domains much smaller than the focal spot. Sample $1L_{S2}$ exhibited a nearly uniform orientation (Figure S8a) but significantly reduced SHG intensity (~15% of the reference). Unlike the samples in Figure 3h and 3l, its SHG image (Figure S8b) lacked visible boundaries, suggesting that the effective domain size is smaller than the probing area (~1 $\mu m^2$). Then, each focal spot contains multiple orientational domains, some of which are antiparallel to others. Interferograms from 25 random spots (Figure S8c) showed that two exhibited a π-phase shift, even though the region appeared uniformly aligned (θ = ~13°). This half-cycle phase difference directly demonstrates that the two spots, each comprising multiple domains, are effectively aligned antiparallel to the others.

### 2.5. Mixed-domain model for SHG suppression

As shown in Figure 2f, the SHG intensity of CVD-grown hBN can be lower than that of mechanically exfoliated hBN by up to three orders of magnitude. Figure 4 further provides direct interferometric evidence for the presence of antiparallel domains. Motivated by these observations, we present a mixed-domain model to quantify how orientational disorder and antiparallel-domain mixing affect the far-field SHG intensity. In the top row of **Figure 5a**, model hBN polycrystals are composed of randomly oriented, equal-sized N domains within a focal area. The FWHM of the orientational distribution (Δθ) was varied from 0° to 30°. Figure 5b shows the normalized $I_{2\omega}^{\parallel}$ calculated in the far field as a function of Δθ. Although the intensity decreased with increasing Δθ, the reduction was modest. For instance, an orientational spread of ~1° observed for $3L_{U1}$ (Figure 5f) resulted in only ~0.1% reduction across different N values—far smaller than the experimentally observed decrease by 500 times (Figure 2f). Even the most disordered sample ($1L_{C1}$) with several-degree spread (Figure 5f) should still yield ~95% of the reference intensity, which is far from ~0.1% (data not shown). These comparisons imply that the reduction in intensity is due to a different type of disorder.

To quantify the effect of antiparallel domains, we modified the mixed-domain model to divide the domains into two groups: UP and DOWN (Figure 5a, bottom row). The DOWN domains, indicated by red arrows, constituted a fraction R and followed a Gaussian distribution centered at 60° (equivalently 180°) with FWHM of Δθ. The remaining UP domains (fraction 1 - R) were centered at 0° with the same Δθ. Figure 5c plots normalized $I_{2\omega}^{\parallel}$ for N = 10, 100 and 1000 as a function of R when Δθ is zero. As R increased, $I_{2\omega}^{\parallel}$ decreased rapidly, reaching 0.1 at R = 0.35 (N = 1000). Reductions by two and three orders of magnitude occurred at R = 0.43 and 0.47, respectively, which aligns well with the experimental observation. Notably, smaller *N* values led to greater fluctuations and larger minima, reflecting statistical imbalance between UP and DOWN populations. In contrast, $I_{2\omega}^{\parallel}$ was given as $(1\text{-}2R)^2$ in the limit as N approaches infinity. Full-scale calculations of $I_{2\omega}^{\parallel}$ also showed a similar trend for finite values of $\Delta\theta$ (Figure S9).

**2.6. SHG-Raman correlated analysis of crystallinity**

We next turn to a cross-sample comparison to translate the above mechanistic insights into practical structural-quality metrics for hBN. As noted in Figure 2f, CVD-hBN samples exhibited much larger intensity variations in SHG than in Raman. To quantify these trends and their correlations, we analyzed four key observables extracted from 10 types of hBN samples (Figure 5d ~ 5f): Raman intensity, SHG intensity, orientational spread and Raman linewidth, the latter of which is known to reflect structural disorder [28]. In Figure 5d, Raman intensity normalized to a monolayer thickness ranged from 25% to 100% of the $1L_{ME}$ reference. The linewidths of CVD-hBN were two to three times larger than those of the reference. Generally, narrower linewidths correlated with higher intensities, though $1L_{P1}$ showed somewhat smaller Raman linewidth relative to its Raman intensity. Because Raman scattering in hBN is nonresonant, Raman intensity can be approximated as proportional to the density ($\rho$) of the intact or slightly disordered unit cells. Thus, reduced Raman intensity implies a larger fraction of severely defective cells residing at the boundaries between antiparallel domains and not contributing to Raman signals. Then, the broadened linewidth can be attributed to spatial inhomogeneity arising from structural disorder near the boundaries or phonon confinement [63]. As domain boundaries have finite width, the spatial extent of boundary regions increases as domain size decreases, which agrees with the intensity-linewidth relation in Figure 5d. This reasoning also suggests that $1L_{P1}$ has disorders that disproportionately reduce Raman intensity without strongly affecting frequency.

The average SHG intensity (Figure 5e) also displayed an inverse relation with linewidth, indicating a general proportionality between Raman and SHG intensities. This is expected since Raman intensity scales with $\rho$, while SHG intensity scales with $\rho^2$ due to its coherent nature. Based on the spread in Raman intensity in Figure 5d, SHG intensity would be expected to vary between ~6% and 100%. However, the observed variation spanned three orders of magnitude. As shown by the mixed-domain analysis (Figure 5a ~ 5c) and the interferometric phase measurements (Figure 4), this discrepancy arises from the presence of antiparallel domains. Figure 5f shows that the orientational spread (e.g., Figure 3g), corrected for measurement noise (Figure S3), positively correlates with the Raman linewidth. While the linewidth varied by only a factor of 3, the orientational spread spanned nearly two orders of magnitude, from 0.1° in ME-hBN to ~5° in commercial CVD-hBN ($1L_{C1}$). This indicates that the average $\overrightarrow{AC}$ direction of each sampled area (~1 μm$^2$) in $1L_{C1}$ is randomly disoriented by several degrees. From Figure 5, we conclude that these four observables provide quantitative indicators of hBN crystallinity, with SHG intensity and orientational spread being even more sensitive to structural disorder. Together with the interferometric SHG results in Figure 4, these metrics provide a practical framework for identifying antiparallel-domain disorder and benchmarking structural quality in continuous CVD-grown hBN, beyond what Raman spectroscopy alone can resolve.

### 2.7. Optical quality metrics for synthesized hBN

The presence of antiparallel domains in continuously grown hBN is not unexpected [64], because growth generally initiates at multiple nucleation sites on catalyst surfaces without orientational preference. Nevertheless, reports of antiparallel domains in continuous films have been rare, most likely due to the technical challenges of their detection. Although high-resolution TEM and STM provide direct structural information, they are impractical at the relevant length scales. For growth on vicinal facets of Cu [17] or Ge [27] intended to enforce unidirectional alignment, the authors needed to monitor domain orientations in the sub-monolayer regime before the completion of continuous films. In this context, SHG spectroscopy and interferometry, combined with Raman spectroscopy, provide a reliable and scalable structural probe as follows.

First, the intensity and linewidth of the Raman $E_{2g}$ mode are straightforward indicators of crystalline quality. As shown in Figure 5d, CVD-grown samples varied from 25–100% in intensity and 100–300% in linewidth relative to references. Whereas linewidth alone reflects disorder, the intensity requires normalization to a reference such as ME-hBN. When substrates are optically equivalent, intensity differences as small as 10% can be resolved. Second, orientational SHG imaging is effective in visualizing crystalline domains and orientational disorder. The spread in orientations within a single domain correlates with structural disorder. Combined with interferometry, SHG polarimetry can identify crystalline domains, including antiparallel domains, as shown here. This approach is most effective when the characteristic domain size is several times larger than the focal spot, but remains informative for smaller domains when the net SHG phase and intensity are interpreted together. Finally, SHG intensity itself serves as a sensitive metric for antiparallel domain size: as domains shrink further below the focal spot, the UP or DOWN fraction approaches R = 0.5 statistically, leading to further reduction in intensity (Figure 5c).

## 3. Conclusions

We reported interferometric SHG spectroscopy capable of differentiating antiparallel domains in 2D hBN, a structural feature that has long eluded optical characterization. By employing a dual-polarization polarimetric scheme, we achieved orientational imaging over macroscopic areas inaccessible to conventional atomic-resolution techniques. Across multiple distinct CVD growth methods, we found that antiparallel domains and their boundaries are pervasive even in films that appear unidirectionally aligned on high-symmetry catalytic substrates. We further established a quantitative framework for structural assessment by correlating SHG intensity with Raman signal strength, spectral linewidth and orientational disorder. Numerical simulations based on a mixed-domain model revealed that the three-orders-of-magnitude variation in SHG intensity is governed by the destructive interference among antiparallel domains and scales with their characteristic size. Altogether, this interferometric SHG approach provides a high-throughput, wide-area probe for evaluating crystalline quality. The methodology is broadly applicable to other non-centrosymmetric 2D materials, including TMDs, offering a powerful platform for guiding growth optimization and domain engineering.

## 4. Methods

### 4.1. Preparation and characterization of samples

Ten different CVD methods were employed to synthesize 2D hBN samples with distinct structural characteristics, as summarized in Table S1. All methods produced continuous films, either monolayers or a few layers. Monolayer hBN films from a commercial source (2D Semiconductors) were included. High-quality n-layer reference samples ($nL_{ME}$) were prepared by mechanical exfoliation [58] of bulk crystals synthesized under high-pressure and high-temperature conditions [49]. Optical excitation and collection from samples supported on $SiO_2$/Si substrates are modulated by multiple reflections and thus depend on $SiO_2$ thickness [65]. For quantitative comparison of optical signals, all samples, unless noted otherwise, were transferred or exfoliated on Si substrates with 85-nm $SiO_2$ epilayers.

### 4.2. Topographic measurements and thickness determination

Topographic characterization was carried out using a commercial AFM unit (Park Systems, XE-70). Height and phase images were obtained in noncontact or tapping mode using Si tips with a nominal radius of 8 nm (MicroMasch, NSC-15). Because of their small thickness, AFM height measurements were not always reliable for determining the number of layers. For CVD-grown samples, thicknesses were determined from cross-sectional electron microscopy. For mechanically exfoliated references, thickness was determined by Si-referenced Raman signals, after pre-screening by optical contrast and SHG intensity (Figure 2e).

### 4.3. Optical contrast measurements

OC was calculated from optical micrographs with RGB color channels: OC = $(R - R_0)/R_0$, where R and $R_0$ are the average red-channel intensities of the sample and bare substrate regions, respectively. While other color channels showed similar trends, the red channel provided the largest contrast values (Figure S10).

### 4.4. Raman measurements

Raman spectra were obtained using a home-built micro-Raman setup described elsewhere [66]. A plane-polarized 514 nm excitation beam was focused onto the sample through an objective lens (40×, numerical aperture = 0.60), producing a focal spot of ~1.0 μm. Back-scattered Raman signals were collected with the same objective and directed to a spectrograph (Princeton Instruments, SP2300) combined with a liquid nitrogen-cooled CCD (Princeton Instruments, PyLon). Excitation power was maintained below 10 mW.

### 4.5. SHG spectroscopy and interferometry

SHG measurements were performed with a home-built micro-SHG spectroscopy setup configured upon a commercial microscope (Nikon, Ti-U) as described elsewhere [33]. The fundamental excitation was provided by a plane-polarized tunable Ti:sapphire laser (Coherent Inc., Chameleon), delivering 140 fs pulses at 80 MHz. The beam was focused onto the samples using microscope objectives (40×, numerical aperture = 0.60; 100×, numerical aperture = 0.90), yielding effective focal spots with FWHMs of 1.3 ± 0.2 μm (40×) and 0.48 ± 0.8 μm (100×). Back-scattered SHG signals were collected with the same objective and guided to a spectrometer equipped with a thermoelectrically cooled CCD (Andor Inc., DU971P). To vary the azimuthal orientation, samples were rotated with a precision of 0.2° about the surface normal using a rotational stage, or the polarization of the excitation beam was varied with a half-wave plate. A linear polarizer was placed before the spectrometer to select the polarization component parallel to that of the incident fundamental beam. For dual-polarization measurements, the polarizer was replaced with a Wollaston prism, enabling simultaneous detection of two orthogonal polarization components.

The in-line spectral phase interferometry based on the above SHG spectroscopy setup was described elsewhere [33]. Briefly, LO SHG pulses ($2\omega_{LO}$) were generated by focusing the fundamental beam at a 100 μm-thick z-cut α-quartz crystal (Figure 4a), which could be rotated to vary the polarization of $2\omega_{LO}$. For the interferometric measurements, the linear polarizer was also used to select the parallel component of the SHG signals. To avoid excessive time delay between LO and sample SHG pulses, a Cassegrain-type reflective objective (52×, numerical aperture = 0.65) was used. The time delay was maintained in the range of 1–3 ps. Spectral interferograms were recorded with the spectrometer and CCD. A grating with 1800 lines/mm was used to achieve optimal spectral resolution. SHG phase was determined using the Fourier transform and filtering [33].

### 4.6. Statistical analysis

For the comparative analysis in Figure 5d ~ 5f, measurements for each modality (SHG and Raman) were performed under identical experimental conditions. Parameters were extracted from randomly sampled spots or regions within each sample. The number of sampled points, varying by sample and parameter, typically exceeded 10. Error bars represent the standard deviation across sampled points, reflecting intra-sample spatial variation.

**Acknowledgements**

We thank Seungil Ahn for the assistance with AFM measurements. This work was supported by the National Research Foundation of Korea (NRF-RS-2024-00336324, NRF-RS-2024-00411134, NRF-2021R1A6A1A10042944) and Samsung Electronics Co., Ltd (IO201215-08191-01). K.K.K. acknowledges support from the Basic Science Research through the National Research Foundation of Korea (NRF), which was funded by the Ministry of Science, ICT and Future Planning, and the Korean government (MSIT) (2022R1A2C2091475 and RS-2024-00439520). H.A. acknowledges support from JSPS KAKENHI (JP24H00407, JP21H05232, JP21H05233).

**Data Availability Statement**

The data supporting the findings of this study are available in the article and its Supplementary Information files.

## References

1. K. S. Novoselov, A. K. Geim, S. V. Morozov, D. Jiang, Y. Zhang, S. V. Dubonos, I. V. Grigorieva, and A. A. Firsov, "Electric Field Effect in Atomically Thin Carbon Films," *Science* 306 (2004): 666.
2. X. S. Li, W. W. Cai, J. H. An, S. Kim, J. Nah, D. X. Yang, R. Piner, A. Velamakanni, I. Jung, E. Tutuc, S. K. Banerjee, L. Colombo, and R. S. Ruoff, "Large-Area Synthesis of High-Quality and Uniform Graphene Films on Copper Foils," *Science* 324 (2009): 1312.
3. Y. Zhan, Z. Liu, S. Najmaei, P. M. Ajayan, and J. Lou, "Large-Area Vapor-Phase Growth and Characterization of Mos2 Atomic Layers on a Sio2 Substrate," *Small* 8 (2012): 966.
4. Y. Stehle, H. M. Meyer, III, R. R. Unocic, M. Kidder, G. Polizos, P. G. Datskos, R. Jackson, S. N. Smirnov, and I. V. Vlassiouk, "Synthesis of Hexagonal Boron Nitride Monolayer: Control of Nucleation and Crystal Morphology," *Chem. Mater.* 27 (2015): 8041.
5. G. Cassabois, P. Valvin, and B. Gil, "Hexagonal Boron Nitride Is an Indirect Bandgap Semiconductor," *Nat. Photonics* 10 (2016): 262.
6. C. Elias, P. Valvin, T. Pelini, A. Summerfield, C. J. Mellor, T. S. Cheng, L. Eaves, C. T. Foxon, P. H. Beton, S. V. Novikov, B. Gil, and G. Cassabois, "Direct Band-Gap Crossover in Epitaxial Monolayer Boron Nitride," *Nat. Commun.* 10 (2019): 2639.
7. Q. Cai, D. Scullion, W. Gan, A. Falin, S. Zhang, K. Watanabe, T. Taniguchi, Y. Chen, E. J. G. Santos, and L. H. Li, "High Thermal Conductivity of High-Quality Monolayer Boron Nitride and Its Thermal Expansion," *Sci. Adv.* 5 (2019): eaav0129.
8. L. H. Li, J. Cervenka, K. Watanabe, T. Taniguchi, and Y. Chen, "Strong Oxidation Resistance of Atomically Thin Boron Nitride Nanosheets," *ACS Nano* 8 (2014): 1457.
9. A. Falin, Q. Cai, E. J. G. Santos, D. Scullion, D. Qian, R. Zhang, Z. Yang, S. Huang, K. Watanabe, T. Taniguchi, M. R. Barnett, Y. Chen, R. S. Ruoff, and L. H. Li, "Mechanical Properties of Atomically Thin Boron Nitride and the Role of Interlayer Interactions," *Nat. Commun.* 8 (2017): 15815.
10. C. R. Dean, A. F. Young, I. Meric, C. Lee, L. Wang, S. Sorgenfrei, K. Watanabe, T. Taniguchi, P. Kim, K. L. Shepard, and J. Hone, "Boron Nitride Substrates for High-Quality Graphene Electronics," *Nat. Nanotechnol.* 5 (2010): 722.
11. L. Wang, I. Meric, P. Y. Huang, Q. Gao, Y. Gao, H. Tran, T. Taniguchi, K. Watanabe, L. M. Campos, D. A. Muller, J. Guo, P. Kim, J. Hone, K. L. Shepard, and C. R. Dean, "One-Dimensional Electrical Contact to a Two-Dimensional Material," *Science* 342 (2013): 614.
12. T. T. Tran, K. Bray, M. J. Ford, M. Toth, and I. Aharonovich, "Quantum Emission from Hexagonal Boron Nitride Monolayers," *Nat. Nanotechnol.* 11 (2016): 37.
13. G. Babu, A. Sawas, N. K. Thangavel, and L. M. R. Arava, "Two-Dimensional Material-Reinforced Separator for Li–Sulfur Battery," *J. Phys. Chem. C* 122 (2018): 10765.
14. L. Sun, G. Yuan, L. Gao, J. Yang, M. Chhowalla, M. H. Gharahcheshmeh, K. K. Gleason, Y. S. Choi, B. H. Hong, and Z. Liu, "Chemical Vapour Deposition," *Nat. Rev. Methods Primers* 1 (2021): 5.
15. X. Xu, Z. Zhang, J. Dong, D. Yi, J. Niu, M. Wu, L. Lin, R. Yin, M. Li, J. Zhou, S. Wang, J. Sun, X. Duan, P. Gao, Y. Jiang, X. Wu, H. Peng, R. S. Ruoff, Z. Liu, D. Yu, E. Wang, F. Ding, and K. Liu, "Ultrafast Epitaxial Growth of Metre-Sized Single-Crystal Graphene on Industrial Cu Foil," *Sci. Bull.* 62 (2017): 1074.
16. L. Song, L. Ci, H. Lu, P. B. Sorokin, C. Jin, J. Ni, A. G. Kvashnin, D. G. Kvashnin, J. Lou, B. I. Yakobson, and P. M. Ajayan, "Large Scale Growth and Characterization of Atomic Hexagonal Boron Nitride Layers," *Nano Lett.* 10 (2010): 3209.

17. L. Wang, X. Xu, L. Zhang, R. Qiao, M. Wu, Z. Wang, S. Zhang, J. Liang, Z. Zhang, Z. Zhang, W. Chen, X. Xie, J. Zong, Y. Shan, Y. Guo, M. Willinger, H. Wu, Q. Li, W. Wang, P. Gao, S. Wu, Y. Zhang, Y. Jiang, D. Yu, E. Wang, X. Bai, Z.-J. Wang, F. Ding, and K. Liu, "Epitaxial Growth of a 100-Square-Centimetre Single-Crystal Hexagonal Boron Nitride Monolayer on Copper," *Nature* 570 (2019): 91.
18. T.-A. Chen, C.-P. Chuu, C.-C. Tseng, C.-K. Wen, H. S. P. Wong, S. Pan, R. Li, T.-A. Chao, W.-C. Chueh, Y. Zhang, Q. Fu, B. I. Yakobson, W.-H. Chang, and L.-J. Li, "Wafer-Scale Single-Crystal Hexagonal Boron Nitride Monolayers on Cu (111)," *Nature* 579 (2020): 219.
19. Y. Shi, C. Hamsen, X. Jia, K. K. Kim, A. Reina, M. Hofmann, A. L. Hsu, K. Zhang, H. Li, Z.-Y. Juang, M. S. Dresselhaus, L.-J. Li, and J. Kong, "Synthesis of Few-Layer Hexagonal Boron Nitride Thin Film by Chemical Vapor Deposition," *Nano Lett.* 10 (2010): 4134.
20. K. Y. Ma, L. Zhang, S. Jin, Y. Wang, S. I. Yoon, H. Hwang, J. Oh, D. S. Jeong, M. Wang, S. Chatterjee, G. Kim, A. R. Jang, J. Yang, S. Ryu, H. Y. Jeong, R. S. Ruoff, M. Chhowalla, F. Ding, and H. S. Shin, "Epitaxial Single-Crystal Hexagonal Boron Nitride Multilayers on Ni (111)," *Nature* 606 (2022): 88.
21. S. Fukamachi, P. Solís-Fernández, K. Kawahara, D. Tanaka, T. Otake, Y.-C. Lin, K. Suenaga, and H. Ago, "Large-Area Synthesis and Transfer of Multilayer Hexagonal Boron Nitride for Enhanced Graphene Device Arrays," *Nat. Electron.* 6 (2023): 126.
22. J. S. Lee, S. H. Choi, S. J. Yun, Y. I. Kim, S. Boandoh, J.-H. Park, B. G. Shin, H. Ko, S. H. Lee, Y.-M. Kim, Y. H. Lee, K. K. Kim, and S. M. Kim, "Wafer-Scale Single-Crystal Hexagonal Boron Nitride Film Via Self-Collimated Grain Formation," *Science* 362 (2018): 817.
23. M. T. Paffett, R. J. Simonson, P. Papin, and R. T. Paine, "Borazine Adsorption and Decomposition at Pt(111) and Ru(001) Surfaces," *Surf. Sci.* 232 (1990): 286.
24. C. M. Orofeo, S. Suzuki, H. Kageshima, and H. Hibino, "Growth and Low-Energy Electron Microscopy Characterization of Monolayer Hexagonal Boron Nitride on Epitaxial Cobalt," *Nano Res.* 6 (2013): 335.
25. P. Sutter, J. Lahiri, P. Albrecht, and E. Sutter, "Chemical Vapor Deposition and Etching of High-Quality Monolayer Hexagonal Boron Nitride Films," *ACS Nano* 5 (2011): 7303.
26. H. Ahn, G. Moon, H.-g. Jung, B. Deng, D.-H. Yang, S. Yang, C. Han, H. Cho, Y. Yeo, C.-J. Kim, C.-H. Yang, J. Kim, S.-Y. Choi, H. Park, J. Jeon, J.-H. Park, and M.-H. Jo, "Integrated 1d Epitaxial Mirror Twin Boundaries for Ultrascaled 2d Mos2 Field-Effect Transistors," *Nat. Nanotechnol.* 19 (2024): 955.
27. J.-H. Jung, C. Zhao, S.-J. Yang, J.-H. Park, W.-J. Lee, S.-B. Song, J. Kim, C.-C. Hwang, S.-H. Baek, F. Ding, and C.-J. Kim, "Step-Directed Epitaxy of Unidirectional Hexagonal Boron Nitride on Vicinal Ge(110)," *Small Struct.* 5 (2024): 2400297.
28. L. Schué, I. Stenger, F. Fossard, A. Loiseau, and J. Barjon, "Characterization Methods Dedicated to Nanometer-Thick hBN Layers," *2D Mater.* 4 (2017): 015028.
29. L. G. Cançado, A. Jorio, E. H. M. Ferreira, F. Stavale, C. A. Achete, R. B. Capaz, M. V. O. Moutinho, A. Lombardo, T. S. Kulmala, and A. C. Ferrari, "Quantifying Defects in Graphene Via Raman Spectroscopy at Different Excitation Energies," *Nano Lett.* 11 (2011): 3190.
30. P. Innocenzi, and L. Stagi, "From Defects to Photoluminescence in h-BN 2D and 0D Nanostructures," *Acc. Mater. Res.* 5 (2024): 413.
31. R. K. Chang, J. Ducuing, and N. Bloembergen, "Relative Phase Measurement between Fundamental and Second-Harmonic Light," *Phys. Rev. Lett.* 15 (1965): 6.
32. K. Kemnitz, K. Bhattacharyya, J. M. Hicks, G. R. Pinto, B. Eisenthal, and T. F. Heinz, "The Phase of Second-Harmonic Light Generated at an Interface and Its Relation to Absolute Molecular Orientation," *Chem. Phys. Lett.* 131 (1986): 285.

33. W. Kim, J. Y. Ahn, J. Oh, J. H. Shim, and S. Ryu, "Second-Harmonic Young's Interference in Atom-Thin Heterocrystals," *Nano Lett.* 20 (2020): 8825.
34. J. Oh, W. Kim, G. Jeong, Y. Lee, J. Kim, H. Kim, H. S. Shin, and S. Ryu, "Dual-Polarization Second Harmonic Generation Interferometry for Imaging Antiparallel Domains and Stacking Angles of 2D Heterocrystals," *ACS Nano* 19 (2025): 37098.
35. Y. Li, Y. Rao, K. F. Mak, Y. You, S. Wang, C. R. Dean, and T. F. Heinz, "Probing Symmetry Properties of Few-Layer $MoS_2$ and h-BN by Optical Second-Harmonic Generation," *Nano Lett.* 13 (2013): 3329.
36. X. Yin, Z. Ye, D. A. Chenet, Y. Ye, K. O'Brien, J. C. Hone, and X. Zhang, "Edge Nonlinear Optics on a $MoS_2$ Atomic Monolayer," *Science* 344 (2014): 488.
37. A. J. Goodman, and W. A. Tisdale, "Enhancement of Second-Order Nonlinear-Optical Signals by Optical Stimulation," *Phys. Rev. Lett.* 114 (2015): 183902.
38. P. E. Ohno, H. Chang, A. P. Spencer, Y. Liu, M. D. Boamah, H.-f. Wang, and F. M. Geiger, "Beyond the Gouy–Chapman Model with Heterodyne-Detected Second Harmonic Generation," *J. Phys. Chem. Lett.* 10 (2019): 2328.
39. W. Kim, G. Jeong, J. Oh, J. Kim, K. Watanabe, T. Taniguchi, and S. Ryu, "Exciton-Sensitized Second-Harmonic Generation in 2d Heterostructures," *ACS Nano* 17 (2023): 20580.
40. E. H. G. Backus, J. Schaefer, and M. Bonn, "Probing the Mineral–Water Interface with Nonlinear Optical Spectroscopy," *Angew. Chem., Int. Ed.* 60 (2021): 10482.
41. Y. Gao, A. J. Goodman, P.-C. Shen, J. Kong, and W. A. Tisdale, "Phase-Modulated Degenerate Parametric Amplification Microscopy," *Nano Lett.* 18 (2018): 5001.
42. N. S. Mueller, A. P. Fellows, B. John, A. E. Naclerio, C. Carbogno, K. Gharagozloo-Hubmann, D. Baláž, R. A. Kowalski, H. H. Heenen, C. Scheurer, K. Reuter, J. D. Caldwell, M. Wolf, P. R. Kidambi, M. Thämer, and A. Paarmann, "Full Crystallographic Imaging of Hexagonal Boron Nitride Monolayers with Phonon-Enhanced Sum-Frequency Microscopy," *Adv. Mater.* 38 (2026): e10124.
43. E. O. Potma, C. L. Evans, and X. S. Xie, "Heterodyne Coherent Anti-Stokes Raman Scattering (Cars) Imaging," *Opt. Lett.* 31 (2006): 241.
44. M. Jurna, J. P. Korterik, C. Otto, J. L. Herek, and H. L. Offerhaus, "Background Free CARS Imaging by Phase Sensitive Heterodyne CARS," *Opt. Express* 16 (2008): 15863.
45. R. W. Boyd, *Nonlinear Optics*, Academic Press, Burlington **2008**.
46. W.-T. Hsu, Z.-A. Zhao, L.-J. Li, C.-H. Chen, M.-H. Chiu, P.-S. Chang, Y.-C. Chou, and W.-H. Chang, "Second Harmonic Generation from Artificially Stacked Transition Metal Dichalcogenide Twisted Bilayers," *ACS Nano* 8 (2014): 2951.
47. C.-J. Kim, L. Brown, M. W. Graham, R. Hovden, R. W. Havener, P. L. McEuen, D. A. Muller, and J. Park, "Stacking Order Dependent Second Harmonic Generation and Topological Defects in H-Bn Bilayers," *Nano Lett.* 13 (2013): 5660.
48. L. Li, Q. Wang, F. Wu, Q. Xu, J. Tian, Z. Huang, Q. Wang, X. Zhao, Q. Zhang, Q. Fan, X. Li, Y. Peng, Y. Zhang, K. Ji, A. Zhi, H. Sun, M. Zhu, J. Zhu, N. Lu, Y. Lu, S. Wang, X. Bai, Y. Xu, W. Yang, N. Li, D. Shi, L. Xian, K. Liu, L. Du, and G. Zhang, "Epitaxy of Wafer-Scale Single-Crystal Mos2 Monolayer Via Buffer Layer Control," *Nat. Commun.* 15 (2024): 1825.
49. K. Watanabe, T. Taniguchi, and H. Kanda, "Direct-Bandgap Properties and Evidence for Ultraviolet Lasing of Hexagonal Boron Nitride Single Crystal," *Nat. Mater.* 3 (2004): 404.
50. S. J. Haigh, A. Gholinia, R. Jalil, S. Romani, L. Britnell, D. C. Elias, K. S. Novoselov, L. A. Ponomarenko, A. K. Geim, and R. Gorbachev, "Cross-Sectional Imaging of Individual Layers and Buried Interfaces of Graphene-Based Heterostructures and Superlattices," *Nat. Mater.* 11 (2012): 764.

51. Y. Ryu, W. Kim, S. Koo, H. Kang, K. Watanabe, T. Taniguchi, and S. Ryu, “Interface-Confined Doubly Anisotropic Oxidation of Two-Dimensional $MoS_2$,” *Nano Lett.* 17 (2017): 7267.
52. R. S. Pease, “An X-Ray Study of Boron Nitride,” *Acta Crystallogr.* 5 (1952): 356.
53. T. Zhang, S. Qiao, H. Xue, Z. Wang, C. Yao, X. Wang, K. Feng, L.-J. Li, and D.-K. Ki, “Accurate Layer-Number Determination of Hexagonal Boron Nitride Using Optical Characterization,” *Nano Lett.* 24 (2024): 14774.
54. A. C. Ferrari, “Raman Spectroscopy of Graphene and Graphite: Disorder, Electron-Phonon Coupling, Doping and Nonadiabatic Effects,” *Solid State Commun.* 143 (2007): 47.
55. C. Lee, H. Yan, L. E. Brus, T. F. Heinz, J. Hone, and S. Ryu, “Anomalous Lattice Vibrations of Single- and Few-Layer $Mos_2$,” *ACS Nano* 4 (2010): 2695.
56. K. F. Mak, M. Y. Sfeir, Y. Wu, C. H. Lui, J. A. Misewich, and T. F. Heinz, “Measurement of the Optical Conductivity of Graphene,” *Phys. Rev. Lett.* 101 (2008): 196405.
57. H. Li, J. Wu, X. Huang, G. Lu, J. Yang, X. Lu, Q. Xiong, and H. Zhang, “Rapid and Reliable Thickness Identification of Two-Dimensional Nanosheets Using Optical Microscopy,” *ACS Nano* 7 (2013): 10344.
58. R. V. Gorbachev, I. Riaz, R. R. Nair, R. Jalil, L. Britnell, B. D. Belle, E. W. Hill, K. S. Novoselov, K. Watanabe, T. Taniguchi, A. K. Geim, and P. Blake, “Hunting for Monolayer Boron Nitride: Optical and Raman Signatures,” *Small* 7 (2011): 465.
59. L. Mennel, M. M. Furchi, S. Wachter, M. Paur, D. K. Polyushkin, and T. Mueller, “Optical Imaging of Strain in Two-Dimensional Crystals,” *Nat. Commun.* 9 (2018): 516.
60. L. Seungjin, C. Soo Ho, K. Hayoung, K. Soo Min, and K. and Ki Kang, “Facile Synthesis of Wafer-Scale Single-Crystal Graphene Film on Atomic Sawtooth Cu Substrate,” *Appl. Sci. Converg. Technol.* 32 (2023): 26.
61. J. I. Dadap, J. Shan, A. S. Weling, J. A. Misewich, and T. F. Heinz, “Homodyne Detection of Second-Harmonic Generation as a Probe of Electric Fields,” *Appl. Phys. B.* 68 (1999): 333.
62. B. C. Bayer, S. Caneva, T. J. Pennycook, J. Kotakoski, C. Mangler, S. Hofmann, and J. C. Meyer, “Introducing Overlapping Grain Boundaries in Chemical Vapor Deposited Hexagonal Boron Nitride Monolayer Films,” *ACS Nano* 11 (2017): 4521.
63. R. J. Nemanich, S. A. Solin, and R. M. Martin, “Light Scattering Study of Boron Nitride Microcrystals,” *Phys. Rev. B* 23 (1981): 6348.
64. X. Song, J. Gao, Y. Nie, T. Gao, J. Sun, D. Ma, Q. Li, Y. Chen, C. Jin, A. Bachmatiuk, M. H. Rümmeli, F. Ding, Y. Zhang, and Z. Liu, “Chemical Vapor Deposition Growth of Large-Scale Hexagonal Boron Nitride with Controllable Orientation,” *Nano Res.* 8 (2015): 3164.
65. D. Yoon, H. Moon, Y. W. Son, J. S. Choi, B. H. Park, Y. H. Cha, Y. D. Kim, and H. Cheong, “Interference Effect on Raman Spectrum of Graphene on Sio2/Si,” *Phys. Rev. B* 80 (2009): 125422.
66. J. E. Lee, G. Ahn, J. Shim, Y. S. Lee, and S. Ryu, “Optical Separation of Mechanical Strain from Charge Doping in Graphene,” *Nat. Commun.* 3 (2012): 1024.

## Figure and captions

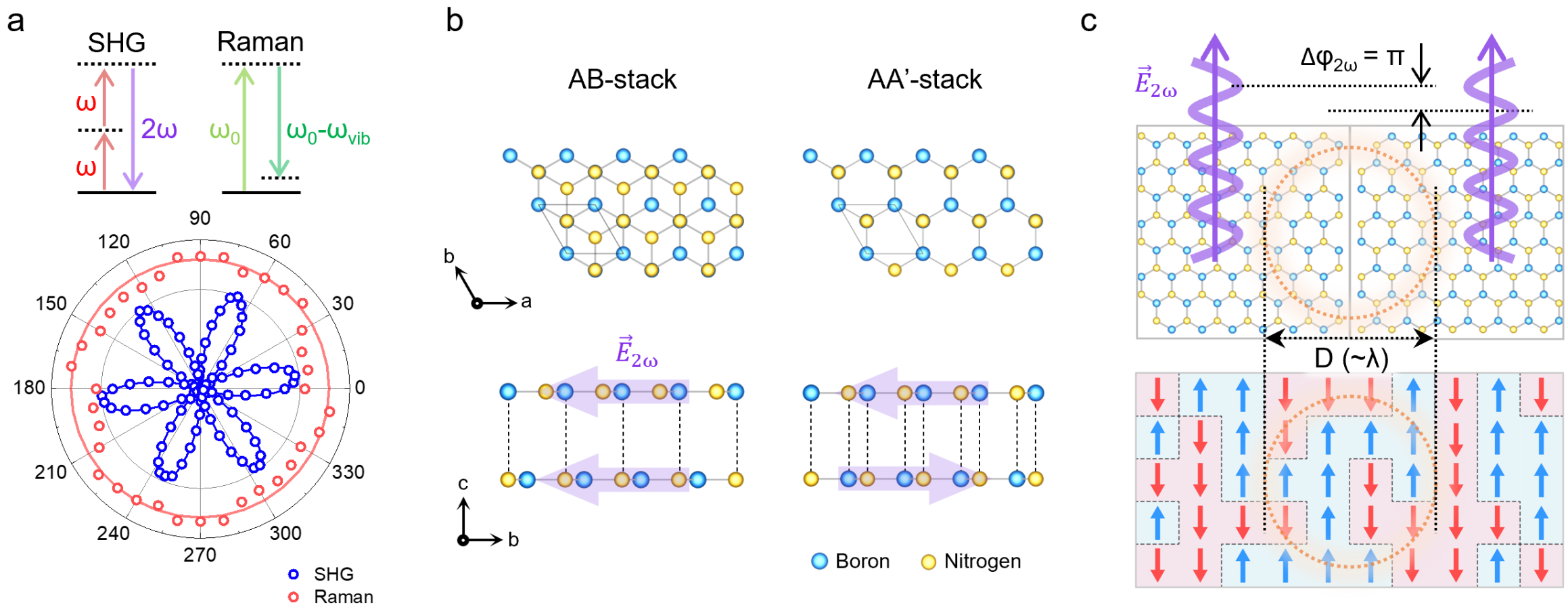

**Figure 1. Structural sensitivity of coherent SHG signals.** (a) Optical processes of SHG and Raman scattering (top) and nonlinear optical anisotropy of 1L hBN (bottom). The polar graphs (bottom) show the orientation dependence of parallel-polarized SHG signals compared with that of Raman scattering. (b) Constructive versus destructive SHG interference in AB- (left) and AA'- (right) stacked 2L hBN. Purple-shade arrows in the side-view schematic (bottom) indicate the directions of SH fields generated in each layer. (c) SH interference in 1L hBN containing antiparallel domains that are larger (top) and smaller (bottom) than the focal spot (diameter D) of the fundamental beam (wavelength λ). Because two antiparallel domains (blue and red arrows in the bottom schematic) generate SHG fields (2ω) with a phase difference ($\Delta\varphi_{2\omega}$) of π, complete destructive interference occurs when the focal spot covers equal areas for the two domain types.

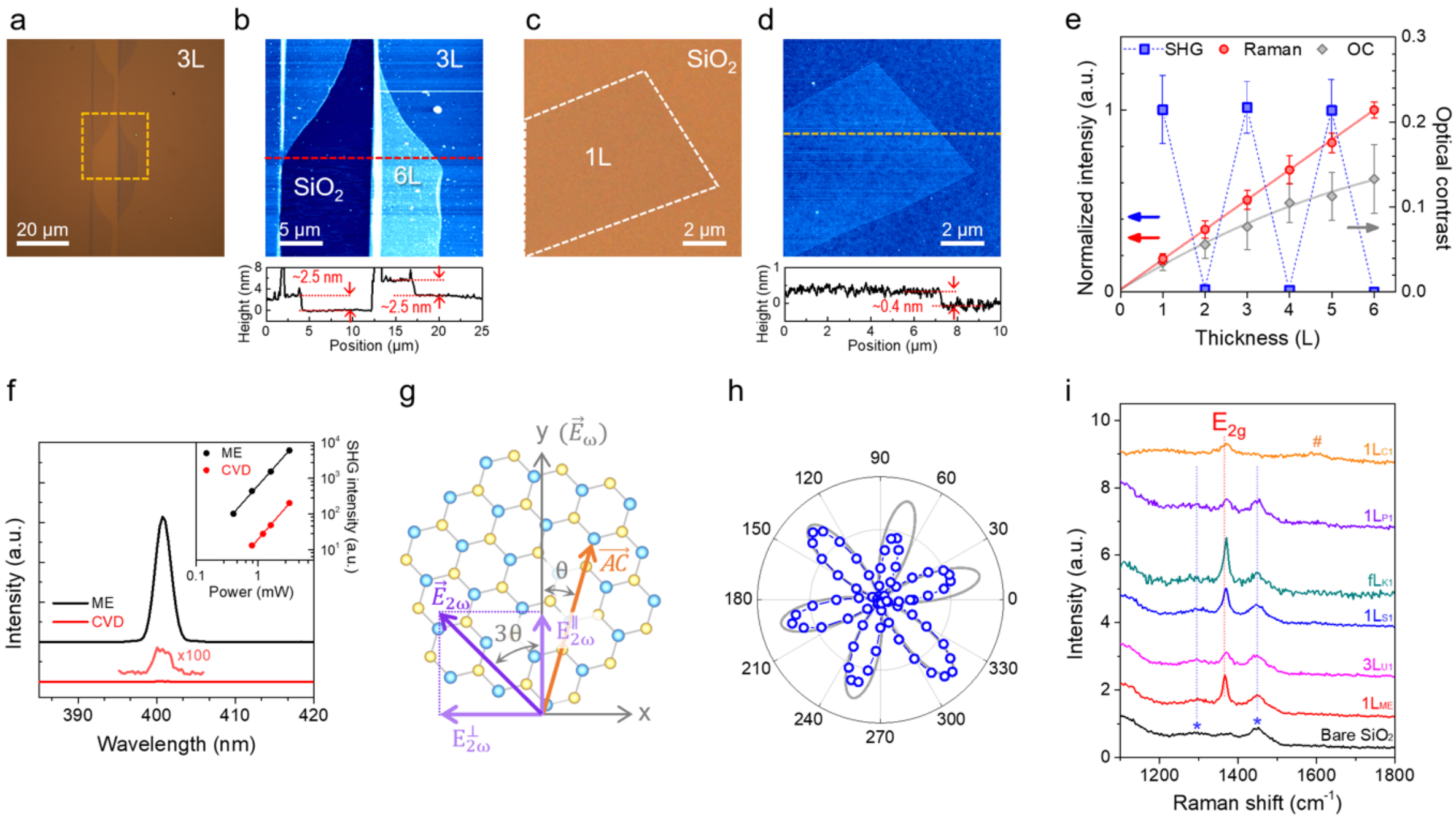

**Figure 2. SHG and Raman characterization of 2D hBN.** (a ~ d) Optical micrographs (a, c) and AFM height images (b, d): CVD-grown 3L hBN ($3L_{U1}$) in (a, b) and mechanically exfoliated 1L hBN ($1L_{ME}$) (c, d). The 1L region in (c) is barely visible and delineated with white dashed lines. A monochrome image in the red-color channel provides enhanced contrast (Figure S10). Height profiles in (b, d) were obtained along the yellow dashed lines in each image. The yellow dashed square in (a) marks the AFM-scanned area shown in (b). (e) SHG (blue squares) and Raman (red circles) intensities as functions of ME-hBN's thickness (left y axis). Optical contrast (gray diamonds) is also given on the right y axis. (f) Representative SHG spectra of $3L_{U1}$ and $1L_{ME}$. Inset: quadratic dependence of SHG intensity on the average power of the fundamental beam. (g) Schematic for the orientation dependence of SH fields ($\vec{E}_{2\omega}$). With θ defined as the angle between the armchair direction ($\overrightarrow{AC}$) and the y axis, $\vec{E}_{2\omega}$ is rotated by 3θ relative to the incident fundamental field ($\vec{E}_{\omega}$), which is aligned along the y axis. Two SH components, $E^{\parallel}_{2\omega}$ and $E^{\perp}_{2\omega}$, were simultaneously detected using the dual-polarization scheme (Methods). (h) Polar graph of parallel SHG intensity ($I^{\parallel}_{2\omega} \propto \left|E^{\parallel}_{2\omega}\right|^2$) from $3L_{U1}$. The solid gray line shows the expected $\cos^2 3\theta$ for ideal 3L hBN. (i) $E_{2g}$ Raman spectra of six representative CVD-hBN samples compared with those of $1L_{ME}$ and bare $SiO_2$/Si substrate. Peaks at 1300 and 1450 $cm^{-1}$ (asterisks) originate from the substrate; the feature near 1590 $cm^{-1}$ for $1L_{C1}$ (sharp symbol) corresponds to the G mode of carbonaceous impurities. $1L_{C1}$ was supported on amorphous quartz, while the other samples were on $SiO_2$/Si substrates.

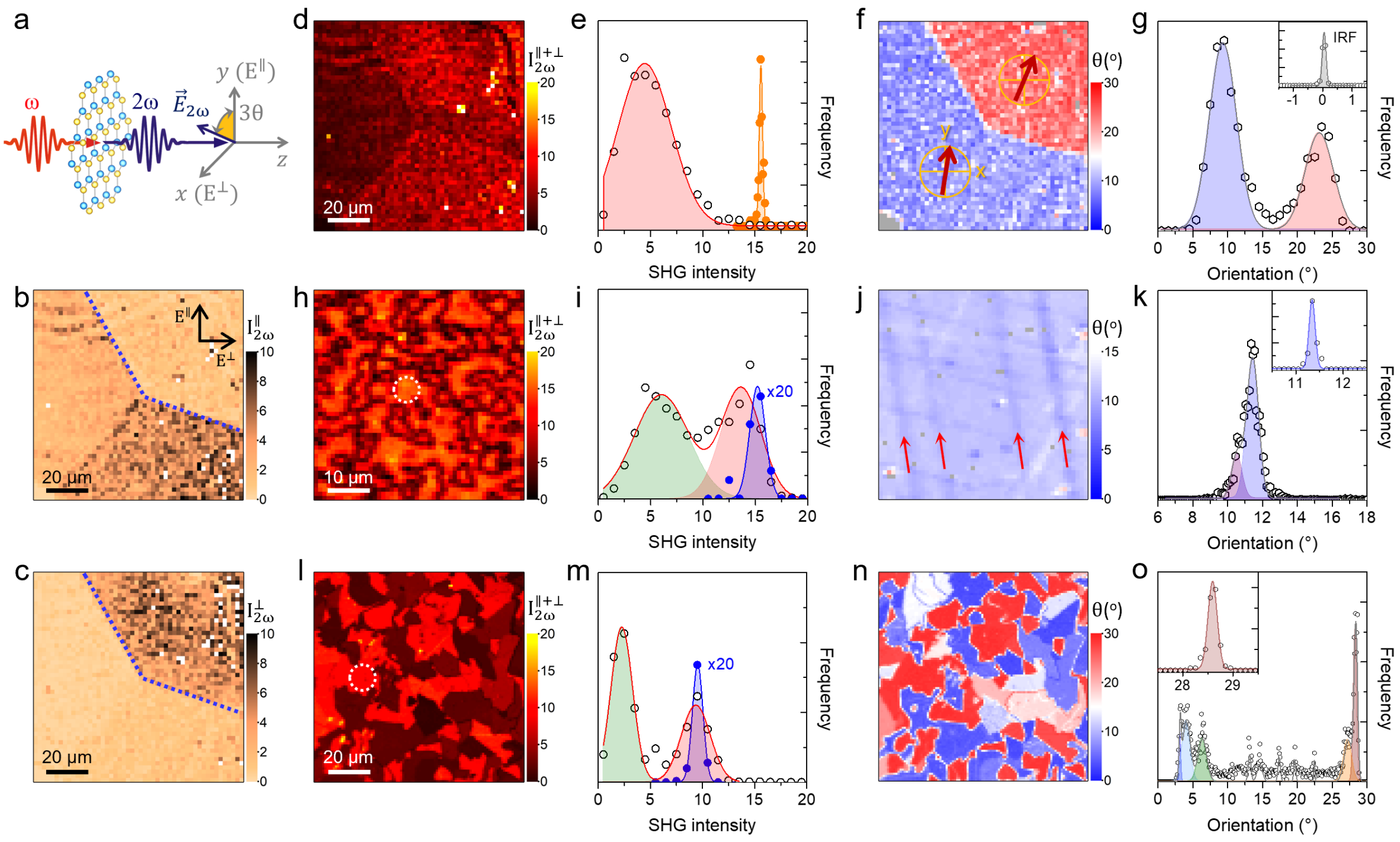

**Figure 3. Orientational SHG imaging of CVD hBN samples.** (a) Schematic of dual-polarization SHG spectroscopy, enabling detection of two orthogonal polarizations ($E_{2\omega}^{\parallel}$ and $E_{2\omega}^{\perp}$). (b, c) Polarized SHG intensity images of $fL_{U3}$: parallel ($I_{2\omega}^{\parallel}$, b) and perpendicular ($I_{2\omega}^{\perp}$, c) to the fundamental polarization aligned along the y axis as depicted in the inset of (b). (d ~ o) Orientational analysis of three types of CVD-hBN samples: $fL_{U3}$ (d ~ g), $1L_{S1}$ (h ~ k), $fL_{K1}$ (l ~ o). Each set includes images of $I_{2\omega}^{\parallel+\perp}$ (d, h, l), frequency graphs of $I_{2\omega}^{\parallel+\perp}$ images (e, i, m), orientation (θ) images (f, j, n) and frequency graphs of θ-images (g, k, o). The brown arrows in (f) denote the average orientation of the two domains (see Figure 3g for the definition of θ). Frequency graphs in (e, i, m) were fitted with Gaussian functions (red and green shades). The blue circles in (i, m) were obtained from the white circles in (h, l), while the orange peak in (e) corresponds to $1L_{ME}$ (Figure S3) shown as a distribution reference. The frequency graphs of θ were also fitted with Gaussian functions: FWHM = 4.6° and 4.7° (g), 0.8° and 0.9° (k), and 0.3–1.6° (o). The inset of (g) shows the frequency graph of θ of $1L_{ME}$, serving as the instrument response function (IRF, FWHM = 0.13°). Insets of (k, o) display the frequency graphs from single domains (white circles in h and l), with FWHM of 0.16° (k) and 0.22° (o).

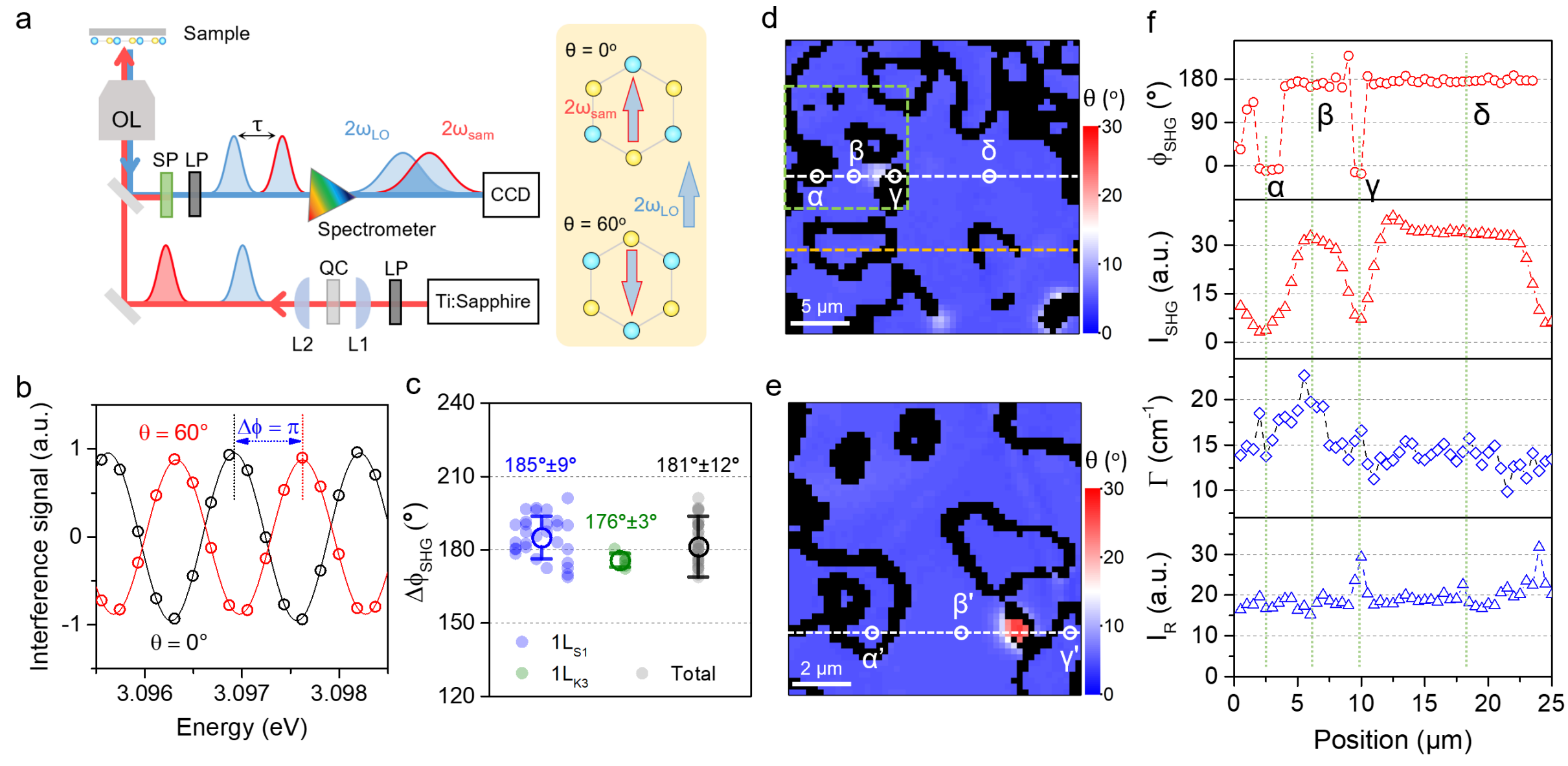

**Figure 4. Interferometric identification of antiparallel domains.** (a) Optical layout of SHG interferometry: convex lenses (L1, L2), α-quartz crystal (QC), objective lens (OL), short-pass filter (SP), linear polarizer (LP), SH pulses of local oscillator and sample ($2\omega_{LO}$ and $2\omega_{sam}$), and delay time (τ). The scheme in the yellow box depicts constructive and destructive interferences between the two pulses. (b) SHG interferograms of $1L_{ME}$ for two antiparallel orientations (θ = 0° and 60°). (c) Cross-domain SHG phase difference ($\Delta\varphi_{SHG}$) of $1L_{S1}$ and $1L_{K3}$. Each value corresponded to crossing two neighboring domains separated by a boundary. A total of 37 $\Delta\varphi_{SHG}$ values, obtained from 8 interferometric line profiles (6 for $1L_{S1}$ and 2 for $1L_{K3}$), clustered near 180°, demonstrating that SHG phase inverts across two antiparallel domains. (d) θ-image of $1L_{S1}$. (e) High-resolution θ-image of $1L_{S1}$, from the green square in (d). α', β', and γ' in (e) correspond to α, β, and γ in (d), respectively. Pixels with SHG intensity < 30% of the average were defined as boundary regions and shown in black (d & e). Unmodified images are given in Figure S6. (f) Line profiles across the white dashed line in (d): (from top to bottom) SHG phase ($\varphi_{SHG}$), SHG intensity ($I_{SHG} = I_{2\omega}^{\parallel+\perp}$), Raman linewidth (Γ) and Raman intensity ($I_R$).

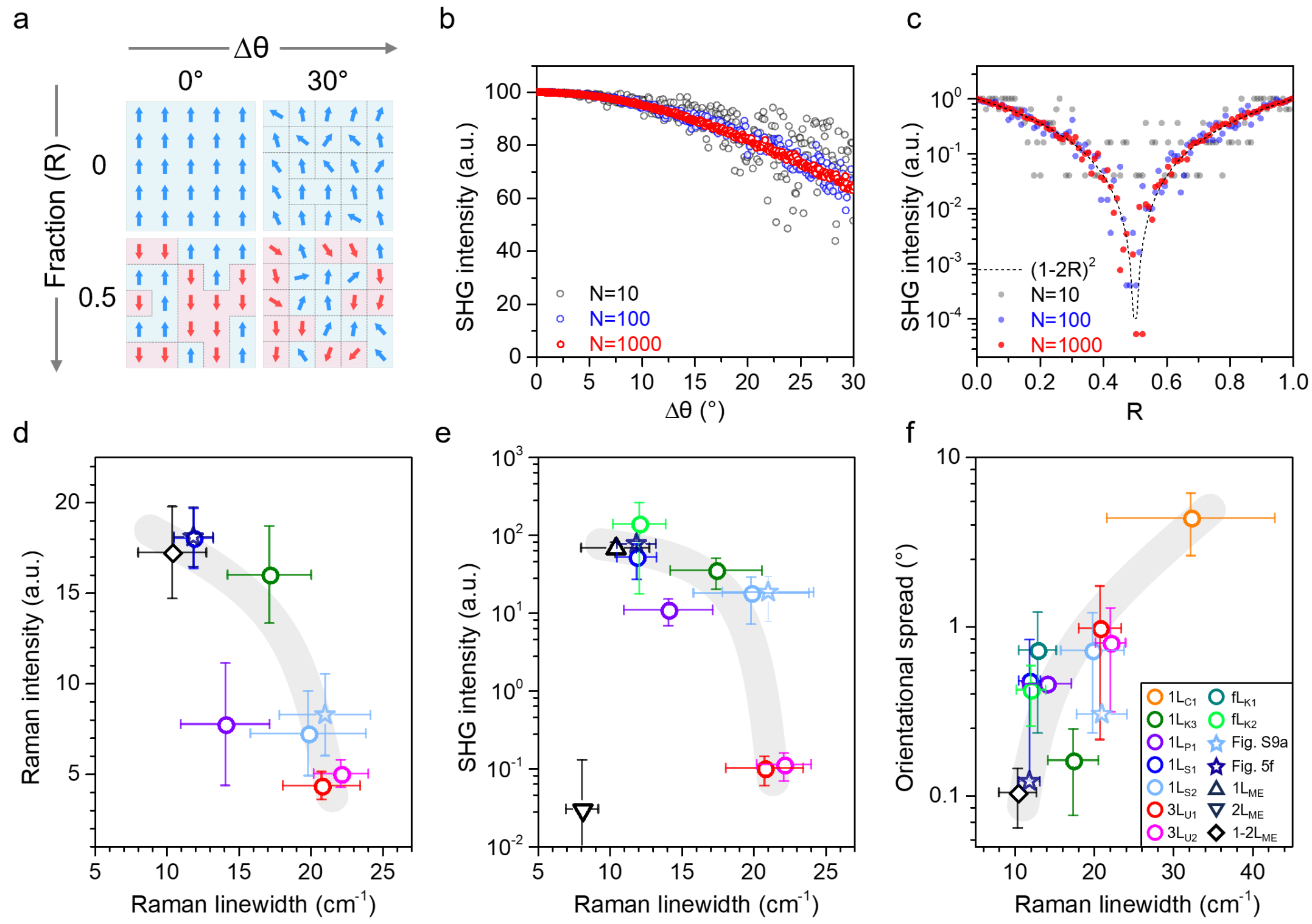

**Figure 5. Mixed-domain model and Raman-SHG correlations for hBN crystallinity.** (a) Mixed-domain model for 1L hBN with UP (blue) and DOWN (red) domains. Parameters: number of domains (N), orientational spread (Δθ) and DOWN-domain fraction (R). (b) Simulated $I_{2\omega}^{\parallel}$ as a function of Δθ for N = 10, 100, and 1000. (c) Simulated $I_{2\omega}^{\parallel}$ as a function of R (Δθ = 0). (d ~ f) Linewidth of $E_{2g}$ Raman mode versus: Raman intensity (a), SHG intensity (b) and orientational spread (c). Raman intensity was normalized by thickness. Shaded lines are drawn as a visual guide.

## Supporting Information

Supporting Information is available from the authors.

## TOC Entry

**Ubiquitous Antiparallel Domains in 2D Hexagonal Boron Nitride Uncovered by Interferometric Nonlinear Optical Imaging**

Interferometric second-harmonic generation (SHG) imaging reveals the ubiquitous formation of antiparallel domains in CVD-grown 2D hexagonal boron nitride. By resolving lattice orientation and SHG phase, hidden domain structures and disorder become visible over large areas. The SHG intensity quantitatively tracks crystalline quality across growth routes, establishing a rapid, non-destructive optical metric for wide-area structural characterization.

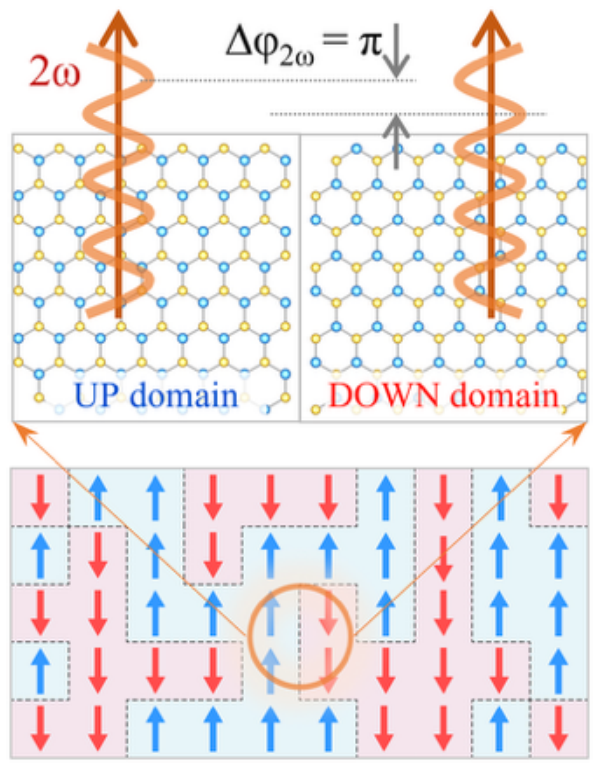